\documentclass{aa}

\usepackage{graphics}

\begin{document}

\thesaurus{20    
  (11.01.2;    
  11.16.1;     
  11.17.3)}    

\title{A search for rapid optical variability in radio-quiet quasars}

\author{M.~Rabbette\inst{1} \and B.~McBreen\inst{1} \and
  N.~Smith\inst{2} \and S.~Steel~\inst{1}} 

\institute{Physics Department, University College, Dublin 4, Ireland
  \and Applied Physics and Instrumentation Department, Regional Technical
  College, Cork, Ireland} 

\date{Received / Accepted} 

\maketitle

\begin{abstract}

The detection of rapid variability on a time\-scale of hours in
radio-quiet quasars (RQQSOs) could be a powerful discriminator between
starburst, accretion disc and relativistic jet models of these
sources. This paper contains an account of a dedicated search for
rapid optical variability in RQQSOs. The technique used differential
photometry between the RQQSO and stars in the same field of view of
the CCD. The 23 RQQSOs that were observed all have high luminosities
($-27<M_\mathrm{V}<-30$), and 22 of these sources are at redshifts 
$z>1$. The total amount of observation time was about 60 hours and
these observations are part of an ongoing programme, started in
September 1990, to search for rapid variability in RQQSOs. No evidence
for short-term variability greater than about 0.1 magnitudes was
detected in any of the 23 sources, however long-term variability was
recorded for the radio-quiet quasar \object{PG 2112$+$059}. The finding charts
are included here because they identify the RQQSO and the reference
stars used in the photometry, and hence are available for use by other
observers.

The unusual properties of two RQQSOs that were not included in our
source list are noted. X-ray results reveal that \object{PG 1416$-$129} is
variable on a timescale of days and that the remarkable source 
\object{IRAS 13349$+$2438} varied by a factor of two on a timescale of a few
hours. The latter source displayed blazar type behaviour in X-rays and
implies that relativistic beaming may occur in at least some RQQSOs.
Radio results also indicate the presence of jets in at least some
RQQSOs.


\end{abstract}

\section{Introduction}

There is a general consensus that quasars belong to two different
radio populations, radio-quiet quasars (RQQSOs) and radio-loud
quasars. $R$ is usually defined as the ratio of the radio
(6~\mbox{cm}) to the optical (440~\mbox{nm}) flux densities and the
radio-quiet quasars have a value of $R<10$, while the radio-loud
quasars have $R>10$ (Kellermann et al. 1989). It is found that
$\sim10~\mbox{\%}$ of quasars are in the radio-loud category. An
additional distinction between active galactic nuclei (AGN) with
strong and weak radio sources comes from the observation that radio
loud objects essentially all occur in elliptical galaxies and RQQSOs
appear to reside in galaxies that are dominated by exponential
disks. However the RQQSOs that occur in elliptical host galaxies are
in general more luminous than those that reside in disks (Taylor et
al. 1996).

Little is known about the short-term variability of ra\-dio-quiet
quasars, because few studies have been carried out (Gopal-Krishna et
al. 1993 and 1995; Jang \& Miller 1995; Sagar et al. 1996). In
contrast blazars display rapid variability in the wavelength range
from radio to gamma rays. The blazar class encompasses both
optically-violent\-ly-variable (OVV) quasars and BL Lac objects and
about one quarter of all radio-loud quasars are also in the blazar
category (Webb et al. 1988; Pica et al. 1988).

There are many theoretical models which endeavour to explain the large
and rapid variability exhibited by blazars and these are usually
divided into extrinsic and intrinsic categories. One extrinsic
mechanism is microlensing of emission knots in a relativistic jet when
they pass behind planets in an intervening galaxy (McBreen \& Metcalfe
1987; Gopal-Krishna \& Subramanian 1991).  The rapid variability from
superluminal-microlensing may be \linebreak responsible for the
variability observed in \object{AO 0235$+$164} (Rabbette et el. 1996)
and \object{PKS 0537$-$441} (Romero et al. 1995). One family of
intrinsic models is based on a rotating supermassive black hole which
accretes matter from a surrounding accretion disc and ejects two
oppositely directed jets. The shocked-jet model involves shocks which
move with relativistic speeds along the jet (Qian et al. 1991;
Marscher, 1980).  It is believed the shock propagates along the line
of sight, through inhomogeneous, small-scale structures distributed
along the jet.  These inhomogeneous structures are illuminated, or
excited, by the moving relativistic shock, through the amplification
of the magnetic field and the acceleration of electrons which causes
the variability in polarization and in flux density that are observed
over a wide range of frequencies (Hughes et al. 1986).  Another family
of intrinsic models invokes numerous flares or hotspots in the
accretion disk and the corona that is believed to surround the central
engine (Wiita et al. 1992; Mangalam \& Wiita 1993) and indeed a
similar model has been proposed to explain X-ray variations in blazars
(Abramowicz et al.  1991). The fact that RQQSOs generally lie on the
far-infrared versus radio correlation (Sopp \& Alexander 1991) suggest
that star formation plays an important role in their radio
emission. It has been suggested by Terlevich et al. (1992) that the
low values of $R$ in RQQSOs can be explained without jets or accretion
discs, by postulating a circumnuclear starburst within a dense,
high-metallicity nuclear environment. In this model the optical/UV and
bolometric luminosity arises from young stars; the variability comes
from cooling instabilities in the shell of compact supernova remnants
and supernova flashes. Variability on intranight timescales is however
difficult to explain with this model because of the short timescales
involved.  Furthermore radio-quiet and radio-loud quasars have very
different radio power outputs but have similar spectral shapes in the
radio region and suggest that a significant fraction of the RQQSOs may
be capable of producing powerful radio emission (Barvainis et
al. 1996).  Kellermann et al. (1994) found possible radio extensions
up to about 300~\mbox{kpc} in a few RQQSOs and assert that for at
least these few cases, the emission is too large to be starburst
related (Stein 1996).

Recently, some evidence suggesting rapid optical variability in the
RQQSOs \object{PG 0946$+$301} and \object{PG 1444$+$407} was reported
by Sagar et al. (1996). They also reported long-term variability for
four RQQSOs. Jang \& Miller (1995) reported intranight variability for
one RQQSO out of a sample of nine sources. Brinkmann et al. (1996)
obtained ASCA observations of the radio-quiet, infrared \linebreak
quasar \object{IRAS 13349$+$2438} and detected substantial X-ray
variability on a timescale of only a few hours.

The results of the photometric observations of a sample of mainly high
luminosity and high redshift RQQSOs are presented. The observations
and data reduction are given in Sect.~2. The results including tables
listing the differential photometry and some light curves are
presented in Sect.~3. The discussion and conclusions are given in
Sects.~4 and 5. Sect.~4 also includes a discussion on two remarkable
RQQSOs, \object{PG 1416$-$129} and \object{IRAS 13349$+$2438}. CCD
images of the fields containing the radio-quiet quas\-ars and reference
stars used in the differential photometry are also included. A value
of $\mathrm{H}_\mathrm{0}=50~\mbox{km s}^{-1}~\mbox{Mpc}^{-1}$ and
$\mathrm{q}_\mathrm{0}=0.5$ has been adopted.

\section{Observations and Data Reduction}

The radio-quiet quasars were selected from the catalogues of
V\'{e}ron-Cetty \& V\'{e}ron (1985), Hewitt \& Burbidge (1987) and
Irwin et al. (1991).  All the sources are at high redshifts ($z>1$)
with the sole exception of \object{PG 2112$+$059} ($z=0.466$) and the
majority of the sources also have high absolute magnitudes
($-27<M_\mathrm{V}<-30$). The photometric data presented here were
obtained during a number of observing runs spanning a period of six
years. The observations were carried out during September 1990,
October/November 1991, August/September 1992, February 1993, March
1994, December 1995 and May/June 1996. All the RQQSOs listed in
Tables~\ref{three-quasars} and \ref{twenty-quasars}, were monitored
during at least one of these observing periods and a few were observed
over a number of years.

The observations were made with the one metre JK Telescope at the
Observatorio del Roque de los Muchachos. A GEC CCD with
$380\times580$ pixels was used for the September 1990 and
October/November 1991 runs. An EEV CCD with $1246\times1152$ pixels
and an image scale of 0.30 arcseconds pixel$^{-1}$ was used for all
the subsequent observing runs. The latter CCD was preferred because
the larger field of view offered a greater choice of reference stars
for the differential photometry. The seeing typically varied between
1.0 and 1.8 arcseconds.

The CCD fields containing the sources were observed through B, V or
R-band filters and integrations times varied between 3 and 14 minutes.
The simultaneous observations of the source and several comparison
stars allowed variations arising from fluctuations in atmospheric
transmission and extinction to be removed. The radio-quiet quasars and
the reference stars used in the analysis are identified in the CCD
images (Fig.~\ref{charts}). Where possible, reference stars of
comparable brightness and colour to the source were selected for the
differential photometry. In all cases two or more reference stars were
used. The CCD frames were processed using the standard software
package IRAF. The DAOPHOT routine was used to carry out
aperture-photometry on each star in the CCD frame. The differential
magnitudes were then calculated for any pair of objects in the frame.

\section{Results}

The dates of observations and the differential photometry results for
each source are presented in Tables~\ref{three-quasars} and
\ref{twenty-quasars}. The source name and redshift are given in
columns 1 and 2. It should be noted that the right ascension (RA) and
declination (DEC) of these sources are included with the finding
charts (Fig.~\ref{charts}). The dates when each source was observed and the
number of observations in each night per filter are presented in
columns 3 and 4 respectively. The differential magnitudes ($\Delta$B,
$\Delta$V, $\Delta$R) between the reference stars (R1, R2, R3) and the
radio-quiet quasar (Q) are given in column 5. In cases where there was
more than one observation of a source in a night, the average value of
the differential magnitudes are given in column 5. The photometric
error ($1\sigma$) is given in column 6.

\begin{figure*}
  \centering
  \vspace{22cm}
  \caption{The finding charts for the radio quiet quasars (RQQ)
    and the reference stars (R1, R2, R3) used in the differential
    photometry. Each frame is orientated so north is up and east to
    the left. The length of the bar is one arc minute. B1950
    coordinates have been used.}
  \label{charts}
\end{figure*}

\begin{table*}
  \caption{B and V-band differential photometry results for the three 
    medium redshift radio-quiet quasars.}
  \centering
  \begin{tabular}{l|l|l|ll|lll|lll|l} \hline
    
    & & & \multicolumn{2}{l|}{\bf number of} & 
    \multicolumn{6}{c|} {\bf differential magnitude}   &\\
    
    {\bf object Name} & {\bf $z$} & {\bf dates of} & 
    \multicolumn{2}{l|}{\bf obs.} & 
    \multicolumn{3}{l|}{\bf $\Delta$B} &
    \multicolumn{3}{l|}{\bf $\Delta$V} & {\bf error} \\
    
    & & {\bf obs.} & \multicolumn{2}{l|}{\bf per filter} & 
    {\bf R1$-$Q} & {\bf R2$-$Q} &
    {\bf R3$-$Q} & {\bf R1$-$Q} & {\bf R2$-$Q} & {\bf R3$-$Q} & {\bf
      $1\sigma$} \\ \hline \ 
    
    {\bf \object{PG 0117$+$213}} & 1.493 & 15-09-90 & 13B & 16V & $-0.59$
    &  $-0.13$ &  --- & $-1.23$  &  $-0.52$ &  --- & 0.05 \\
    
    & & 19-09-90 & 14B & 15V & $-0.59$ & $-0.12$ & --- & $-1.22$ &
    $-0.51$ &  --- & 0.05 \\
    
    & & 31-10-91& 4B & 5V & $-0.56$ & $-0.11$ & --- & $-1.23$ &
    $-0.53$ &  --- & 0.02 \\
    
    & & 01-11-91& 5B & 6V & $-0.57$ & $-0.13$ & --- & $-1.23$ &
    $-0.53$ &  --- & 0.02 \\
    
    & & 21-09-92& 1B & 1V & $-0.57$ & --- & 3.10 & $-1.19$ & --- &
    2.28 & 0.04 \\
    
    & & 22-09-92& 1B & 1V & $-0.58$ & --- & 3.21 & $-1.18$ & ---
    &   ---  & 0.04 \\
    
    & & 23-09-92& 1B & 1V & $-0.54$ & --- & 3.21 & $-1.20$ & ---
    &  ---  & 0.04 \\
    
    & & 24-09-92& 1B & 1V & $-0.56$ & --- & 3.17 & $-1.24$ & --- &
    2.30    & 0.04 \\
    
    & & 25-09-92 & --- & 1V & --- & --- & --- & $-1.23$ & --- &
    2.39    & 0.04 \\
    
    & & 26-09-92& 1B & 1V & $-0.55$ & --- & --- & $-1.21$ & --- &
    2.33    & 0.04 \\
    
    & & 27-09-92& 1B & 1V & $-0.57$ & --- & 3.35 & $-1.24$ & --- &
    2.32 & 0.04 \\
    
    {\bf \object{PG 2112$+$059}} & 0.466 & 15-09-90 & --- & 8V & --- &
    --- & --- & 0.35  & 0.66 &  --- & 0.03 \\
    
    & & 16-09-90 & 6B & 6V & 0.76 & 1.34 & --- & 0.35 & 0.65 &
    --- & 0.03 \\
    
    & & 21-09-92 & 2B & 2V & 0.72 & 1.34 & 0.82 & 0.31 & 0.61 &
    0.23 & 0.03 \\
    
    & & 22-09-92 & 1B & 1V & 0.71 & 1.33 & 0.82 & 0.33 & 0.62 &
    0.21 & 0.03 \\
    
    & & 25-09-92 & 2B & 2V & 0.73 & 1.34 & 0.84 & 0.32 & 0.60 &
    0.22 & 0.03 \\

    & & 26-09-92 & 1B &    1V & 0.72  & 1.32 &  0.82 & 0.33 &  0.60 &
    0.23 & 0.03 \\

    & & 28-05-96 & --- & 2V & --- & --- & --- & 0.51 &  0.81 &  0.41 &
    0.03 \\
    
    & & 31-05-96 & --- & 3V & --- & --- & --- & 0.50 & 0.80 & 0.39 &
    0.03 \\

    & & 01-06-96 & --- & 2V & --- & --- & --- & 0.49  & 0.79 & 0.40 &
    0.03 \\

    & & 03-06-96 & --- & 4V & --- & --- & --- & 0.50 &  0.77  & 0.42 &
    0.03 \\

    {\bf \object{PG 2302$+$029}} & 1.044 & 26-10-91 & 1B &    1V & 0.22  &
    0.59 & --- & $-0.44$  & 0.19 & --- & 0.03 \\

    &  &  31-11-91 & 2B &    2V & 0.26  &  0.54  & --- & $-0.40$ &  0.24 &
    --- & 0.03 \\

    &  &  01-11-91 & 1B &    1V & 0.25  &  0.56 & ---&    ---  &  0.22 &
    ---& 0.03 \\

    &  &  02-11-91 & 1B &    1V & 0.25  &  0.60 & ---& $-0.40$ &  0.24 &
    --- &  0.03 \\

    &  &  20-08-92 & 1B &    1V & 0.27  &  0.54 & --- & $-0.38$  &  0.24 &
    --- & 0.03 \\ \hline
    
  \end{tabular}
  \label{three-quasars}
\end{table*}

\begin{figure*}
  \resizebox{\hsize}{!}{\includegraphics{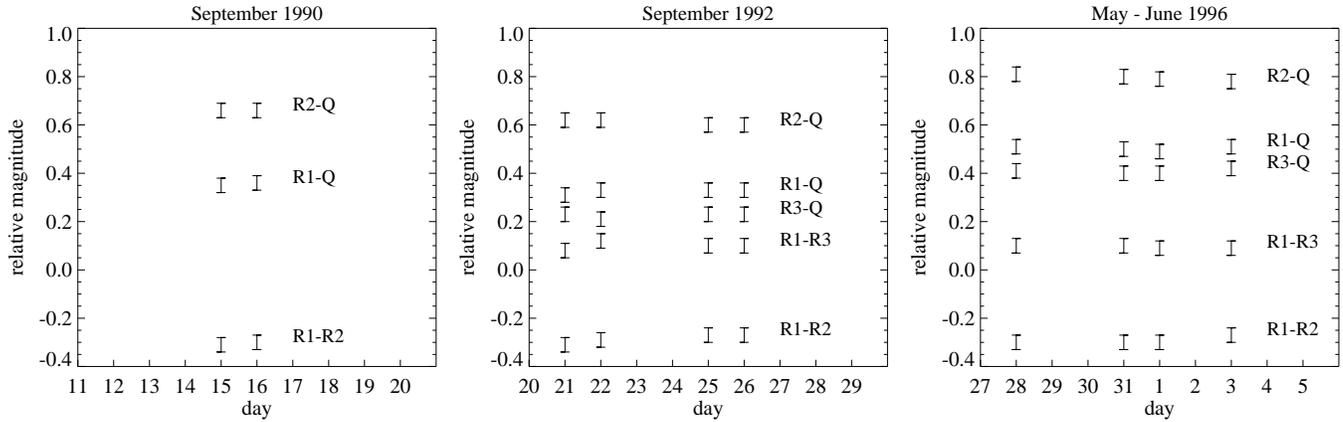}}
  \caption{Differential V band photometry between the radio 
    quiet quasar \object{PG 2112$+$059} (QSO) and three reference stars
    R1, R2 and R3 and differential photometry between the reference
    stars R1-R2 and R1-R3.  The star R3 was unobservable during
    September 1990, due to the small CCD field of view. The average
    relative magnitude per night is plotted.}
  \label{PG2112+059}
\end{figure*}

\begin{table*}
  \caption{V {\it or} R-band differential photometry results for 
    twenty high redshift radio-quiet quasars.}
  \centering
  \begin{tabular}{l|l|l|l|lll|l} \hline
    
    & & & {\bf number of} & & &&   \\

    {\bf object Name}& {\bf $z$}&{\bf dates of}& {\bf observations} &
    \multicolumn{3}{l|}{\bf differential magnitude}  & {\bf error} \\

    & & {\bf observations} & {\bf per filter}&  {\bf  R1$-$Q} & {\bf
    R2$-$Q} & {\bf R3$-$Q}  & {\bf $1\sigma$} \\ \hline
    
    {\bf \object{US 1420}} & 1.473 & 18-02-93 & 2V & $-0.90$ &  $-1.78$ &
    0.30 & 0.04 \\

    & & 20-02-93 & 1V & $-0.94$ &  $-1.80$ &  0.28 & 0.04 \\

    & & 21-02-93 & 2V & $-0.94$ &  $-1.82$ &  0.34 & 0.04 \\

    & & 23-02-93 & 1V & $-0.93$  &   $-1.82$  &   0.32 &  0.04 \\

    {\bf \object{US 1443}} & 1.564 & 18-02-93 & 2V & $-0.97$ &  $-0.74$  &
    $-0.52$ &  0.03 \\

    &  & 21-02-93 & 2V & $-0.95$  &  $-0.75$  &  $-0.50$ &  0.03 \\

    &  & 23-02-93 & 2V & $-0.98$  &  $-0.72$  &  $-0.52$ &  0.03 \\

    {\bf \object{US 1498}} & 1.406 & 18-02-93 & 2V &   0.10  &  $-0.44$
    &  0.37 & 0.03 \\

    & & 21-02-93 & 2V & 0.10  &   $-0.44$ &  0.35 &  0.03 \\

    & & 23-02-93 &  2V & 0.09   &  $-0.45$  & 0.34 & 0.03 \\

    {\bf \object{BR 0945$-$04}} &  4.118  &  22-02-93 & 2R & $-3.59$  &
    $-2.02$ & --- & 0.07 \\

    &  & 24-02-93 & 2R &  $-3.52$ & $-1.95$ & --- & 0.03 \\

    & & 19-12-95 & 3R & $-3.54$ & $-1.97$ & --- & 0.03 \\

    {\bf \object{H 1011$+$091}} & 2.27 & 18-02-93&2R & --- & $-1.15$ &
    $-1.61$ & 0.03 \\

    & & 21-02-93 & 2R & 0.12  & 1.12 & $-1.58$ & 0.03 \\

    & & 23-02-93&2R & 0.11 & $-1.10$ & $-1.60$ & 0.03 \\

    {\bf \object{BRI 1013$+$00}} & 4.38& 22-02-93& 2R & $-2.76$ & $-2.89$ &
    ---  & 0.03 \\ 

    & &24-02-93& 2R & $-2.78$ & $-2.89$ & --- & 0.03 \\

    {\bf \object{BR 1033$-$03}} & 4.50& 22-02-93 & 2R & $-1.13$ & $-2.59$  &
    --- & 0.03 \\

    {\bf \object{BRI 1050$-$00}} & 4.29 & 22-02-93 & 2R & $-0.37$ & 0.20 &
    --- & 0.04 \\

    & & 19-12-95& 2R & $-0.32$  &  0.22 & --- & 0.04 \\

    {\bf \object{BRI 1108$-$07}} & 3.94 & 22-02-93& 2R & $-0.38$  &  $-1.83$ &
    --- & 0.04 \\

    {\bf \object{BRI 1110$+$01}} & 3.93 & 22-02-93& 2R & $-1.67$ & $-1.87$ &
    --- & 0.04 \\

    {\bf \object{BR 1144$-$08}} & 4.16 & 22-02-93 & 2R & $-2.60$ & $-1.83$ &
    --- & 0.04 \\

    {\bf \object{1146$+$111D}} & 2.12 & 18-02-93& 2R & $-0.20$ & 0.15 &  $-2.03$
    & 0.03 \\

    & & 20-02-93 & 2R & $-0.20$ & 0.13 & $-2.04$ & 0.03 \\

    & & 21-02-93& 2R & $-0.18$ & 0.17 & $-2.00$ & 0.03 \\

    & & 23-02-93& 2R & $-0.19$ & 0.18 & $-1.99$ & 0.03 \\

    {\bf \object{1159$+$123}} & 3.51 & 18-02-93 & 2R & 0.23 & $-0.29$ & 0.44 &
    0.04 \\

    & & 20-02-93 & 2R & 0.21 & $-0.28$ & 0.47 & 0.04 \\

    & & 21-02-93 & 2R & 0.20 & $-0.29$ & 0.44 & 0.04 \\

    {\bf \object{1201$-$015}} &2.26& 19-02-93& 2R & $-3.39$ & $-2.38$  & ---
    & 0.04 \\

    & & 21-02-93& 1R & $-3.44$ & $-2.34$ & $-0.68$ & 0.04 \\

    & & 23-02-93& 2R & $-3.45$ & $-2.33$ & $-0.67$ & 0.04 \\

    {\bf \object{BR 1202$-$07}} & 4.70 & 22-02-93 & 2R & $-0.03$ & $-0.61$ &
    --- & 0.07 \\

    {\bf \object{1222$+$023}} & 2.05 & 19-02-93 & 2R & $-3.00$ & 0.92 &
    $-1.81$ & 0.05 \\

    & & 21-02-93& 2R & $-3.02$ & 0.92 & $-1.81$ & 0.05 \\

    & & 23-02-93& 2R & $-2.98$ & 0.93 & $-1.76$ & 0.05 \\

    {\bf \object{PG 1247$+$268}} & 2.041 & 18-02-93& 1R & 1.37 & 2.03 &
    1.73 & 0.03 \\

    & & 21-02-93 & 2R & 1.35 & 2.06 & 1.77 & 0.03 \\

    {\bf \object{W 61972}} & 1.92 & 18-02-93 & 2R & 0.51 & $-1.20$ &
    $-1.83$ & 0.02 \\

    & & 21-02-93 & 2R & 0.51 & $-1.20$ & $-1.82$ & 0.02 \\

    & & 23-02-93& 1R &   ---   & $-1.20$ &  $-1.85$ & 0.02 \\

    {\bf \object{PG 1329$+$412}} & 1.93 & 18-02-93 & 2R & $-0.67$ & 0.79 &
    $-1.55$ & 0.03 \\

    & & 23-02-93 & 2R & $-0.67$ & 0.77 & $-1.54$ & 0.03 \\

    {\bf \object{BRI 1500$+$08}} & 3.96 & 23-02-93 & 2R & $-0.82$ &  $-0.92$
    & --- & 0.07 \\
    
    & & 17-03-94 & 2R & $-0.86$ & --- & --- & 0.07 \\ \hline

  \end{tabular} 
  \label{twenty-quasars}
\end{table*}

\begin{figure*}
  \resizebox{\hsize}{!}{\includegraphics{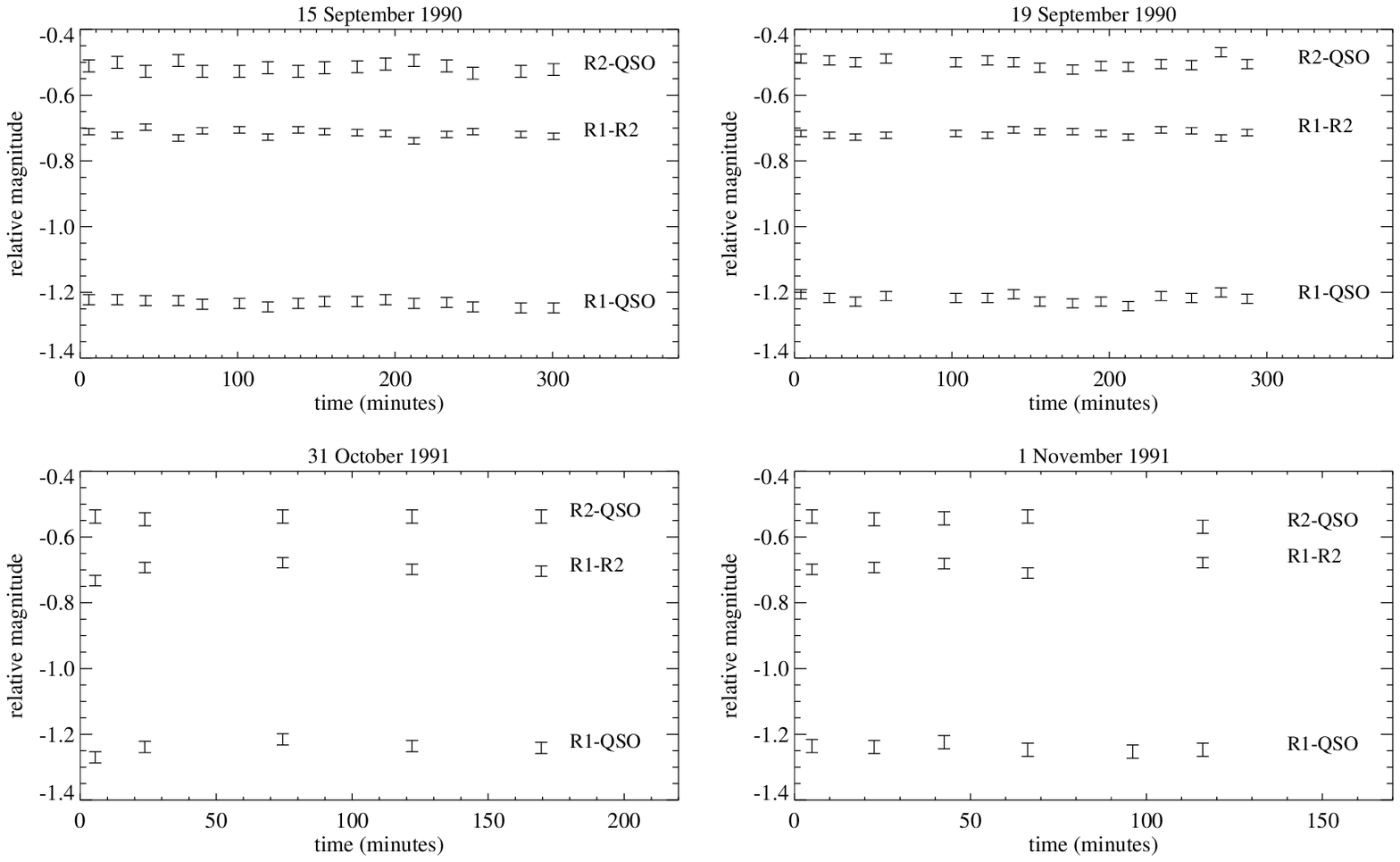}}
  \caption{Differential V band photometry between the radio
    quiet quasar \object{PG 0117$+$213} (QSO) and the two reference
    stars R1 and R2 and differential photometry between the reference
    stars R1-R2.  The origins of the time axes of the two upper plots
    are 00:50 UT on September 15 and 00:32 UT on September 19, 1990.
    The origins of the time axes of the lower plots are 22:12 UT on
    October 31 and 22:43 UT on November 1, 1991.}
  \label{PG0117+213}
\end{figure*}

The major result is that no statistically significant rapid
variability was observed for any of the sources listed
Tables~\ref{three-quasars} and \ref{twenty-quasars}. The data also
reveal no night to night or longer term variability for any of the
sources, with the sole exception of \object{PG 2112$+$059}. This
source brightened by 0.18 magnitudes in the V-band between September
1992 and June 1996 (Table~\ref{three-quasars} and
Fig.~\ref{PG2112+059}). The source with the largest number of
observations is \object{PG 0117$+$213} (Table~\ref{three-quasars}) and
some of the differential V-band results are presented in
Fig.~\ref{PG0117+213}.  No significant short term variability was
observed.  Furthermore careful analysis of the remainder of the V-band
data and all of the B-band revealed no significant short or long term
variability.

The sources in this sample can be divided into three main redshift
groups; A($z<2$), B($2<z<3$), C($z>3$), and the following section
includes comments on a number of the sources in these three
categories.

\begin{description}
\item[\bf A($z<2$)] \object{PG 0117$+$213} was the most frequently
  observed source in the sample and no significant variability was
  detected.  The spectra of the source reveal the emission lines
  \ion{C}{iii} and \ion{C}{iv} at $z_\mathrm{em}=1.493$ (Hewitt \&
  Burbidge 1987).  Infrared observations over a period of years show
  maximum variability of 0.05 and 0.16 magnitudes at 2.2~\mbox{$\mu$m}
  and 10.1~\mbox{$\mu$m} respectively (Neugebauer et al. 1989). Sagar
  et al. (1996) observed this RQQSO during November 1996 and detected
  a hint of optical microvariability and suggested that careful
  monitoring of this source should continue. The only source in this
  sample of RQQSOs to display variability is \object{PG 2112$+$059}.
  It was observed during 1990, 1992 and 1996 and was found to have
  brightened by 0.18 magnitudes between September 1992 and June 1996
  (Table~\ref{three-quasars}). \object{PG 2302$+$029} reveals emission
  peaks in its spectrum which have been identified with \ion{Fe}{iii}
  multiplets (Wampler, 1986). This source has a redshift $z=1.044$ and
  $m_\mathrm{B}=16.03$. No variability was detected during the 1991 or
  1992 observing runs. \object{US 1420}, \object{US 1443} and
  \object{US 1498} are all ultraviolet excess sources.  Spectroscopy
  of blue and ultraviolet excess sources was reported Mitchell et al.
  (1984) and the redshifts are given in Table~\ref{twenty-quasars}.
  These sources and the remaining sources in this category displayed
  no rapid variability.

\item[\bf B($2<z<3$)] \object{H 1011$+$091} is a broad absorption line
  \linebreak (BAL) RQQSO with $m_\mathrm{V}=17.8$ (Hartig \& Baldwin
  1986).  The \ion{Mg}{ii} emission line yields $z=2.27$ (Drew \&
  Boksenberg 1984).  The source \object{1146$+$111D} has a redshift
  $z=2.12$ and is part of a compact group of five quasars with similar
  apparent magnitude and redshift and all within a diameter of 4
  arcminutes (Hazard et al. 1979). The sources \object{1201$-$015} and
  \object{1222$+$023} were detected optically and listed by MacAlpine
  \& Williams (1981). \object{1201$-$015} has an estimated
  $m_\mathrm{V}=18.0$, and emission lines $\lambda$3970 and
  $\lambda$4560 believed to be \ion{Ly}{$\alpha$} and \ion{O}{iv}
  yielding $z=2.26$; \object{1222$+$023} has $z=2.05$ and an estimated
  $m_\mathrm{V}=17.0$. \object{PG 1247$+$268} was observed by Green et
  al. (1980). The optical spectrum shows four strong emission features
  \ion{Ly}{$\alpha$}$+$\ion{N}{v}, \ion{Si}{iv}$+$\ion{O}{iv},
  \ion{C}{iv} and \ion{C}{iii} with $z=2.041$ and $m_\mathrm{V}=15.8$.

\item[\bf C($z>3)$] \object{1159$+$123} discovered by Hazard et al. (1984)
  using objective prism plates from the UK Schmidt telescope is a
  strong emission-line RQQSO with $z_\mathrm{em}=3.51$ and
  $m_\mathrm{V}=17.5$. Irwin et al. (1991) carried out a very
  successful multicolour survey in B, R and I (selected B-R) to search
  for high redshift quasars. Using this method they found 27 quasars
  at $z>4$. R-band data for nine of these high redshift quasars are
  given in Table~\ref{twenty-quasars}: \object{BR 0945$-$0411},
  $m_\mathrm{R}=18.80$; \object{BRI 1013$+$0035}, $m_\mathrm{R}=18.80$;
  \object{BR 1033$-$0327}, $m_\mathrm{R}=18.50$; \object{BRI 1050$-$0000},
  $m_\mathrm{R}=18.59$; \object{BRI 1108$-$0747}, $m_\mathrm{R}=18.13$;
  \object{BRI 1110$+$0106}, $m_\mathrm{R}=18.30$; \object{BR 1144$-$0723},
  $m_\mathrm{R}=18.60$; \object{BR 1202$-$0725}, $m_\mathrm{R}=18.70$
  and \object{BRI 1500$+$0824}, $m_\mathrm{R}=19.25$.  No variability
  was found, however the search for optical variability in these high
  redshift RQQSOs will continue.

\end{description}

\section{Discussion}

The mechanisms that make some quasars radio-loud and others
radio-quiet are not well understood but the major differences may be
attributable to the spin of the black hole (Wilson \& Colbert 1995).
There are a number of well established distinctions between the two
classes. (i) Radio-loud quasars are associated with elliptical host
galaxies and radio-quiet quasars tend to reside in spiral galaxies.
The mean absolute magnitude of the underlying galaxies of radio-loud
quasars is similar to that of radio galaxies, however the host
galaxies of radio-quiet quasars are 0.6-1.0 magnitudes less luminous
(Smith et al. 1986: V\'{e}ron-Cetty \& Woltjer 1990). Recent deep near
infrared (K-band) imaging of host galaxies of quasars revealed that
more than half of RQQSOs appear to lie in galaxies that are dominated
by an exponential disk (Taylor et al.  1996). Those RQQSOs that have
elliptical host galaxies show signs of interaction and are in general
more luminous than those that reside in disk galaxies. There is also
some evidence to suggest that a large majority of low-luminosity
radio-quiet AGN lie in disk galaxies but a significant fraction of
RQQSOs more luminous than $M_\mathrm{V}\approx-23.5$ have elliptical
host galaxies. (ii) Unlike RQQSOs, radio-loud quasars produce large
scale jets and lobes and can be defined by their radio luminosity --
those with
$\mathrm{L}(5~\mbox{GHz})<10^{25}~\mbox{W~Hz}^{-1}~\mbox{sr}^{-1}$ are
classified as RQQSOs, and those with
$\mathrm{L}(5~\mbox{GHz})>10^{25}~\mbox{W~Hz}^{-1}~\mbox{sr}^{-1}$ as
radio loud quasars (Miller et al. 1990). (iii) Gamma-ray results
recently obtained by the Compton Gamma Ray Observatory (CGRO) have
produced evidence for two classes of AGN (Dermer \& Gehrels 1995).
These classes are defined by their redshift, luminosity distributions
and high energy spectral properties. The first class of objects have
redshifts $\leq0.06$ and 50--150~\mbox{keV} luminosities in the range
$10^{41}$--$10^{44}~\mbox{ergs}^{-1}$. Associated with this group are
Seyfert galaxies, RQQSOs and radio-galaxies viewed at large angles
with respect to the radio jet axis. The gamma-ray spectra for these
sources soften between $\sim100~\mbox{keV}$ and several MeV.  The
second class of source consists of blazars with redshifts as large as
2.3 which have detectable fluxes of gamma-rays above 100~\mbox{MeV}
with emission extending into the GeV band. This class of source
probably consists of AGN that are observed close to the axis of a
radio jet. So far no RQQSO has been detected at gamma ray energies
above 10~\mbox{MeV}.

Unlike radio-loud quasars, intensive optical monitoring of RQQSOs only
started in the early 1990s. Gopal-Krishna et al. (1995) monitored a
sample of six optically bright and luminous RQQSOs
($m_\mathrm{V}\approx16$ and $M_\mathrm{V}<-23$) and found strong
hints for variability in three RQQSOs; \object{PG 0946$+$301}
displayed an R-band increase of $\sim0.05$ magnitudes in a time of
$\sim0.5~\mbox{hr}$; \object{PG 1049$-$006} varied in V-band by
$\sim0.05$ magnitudes in $\sim0.6~\mbox{hr}$ and \object{PG 1206$+$459}
varied by 0.04 magnitudes in $\sim2~\mbox{hr}$ in the R-band. Sagar et
al. (1996) reported that the flux density from \object{PG 1444$+$407}
dropped by 0.04 magnitudes in $\sim0.5~\mbox{hr}$ in the R-band. Sagar
et al.  (1996) also reported long-term variability (over
$\sim1~\mbox{yr}$) for four RQQSOs in their sample, with the largest
variability of about 0.15 magnitudes in 11 months, recorded for the
source \object{PG 1049$-$005}. Jang \& Miller (1995) reported
intranight variability in the RQQSO \object{II Zw 175} which varied by
$\sim0.05$ magnitudes over a period of 4~days. The reported detection
of variability in RQQSOs is not in conflict with the results presented
here and in Tables~\ref{three-quasars} and \ref{twenty-quasars},
because the small levels of variability would not have been detected
above the $3\sigma$ level in many of the sources monitored in this
survey. One source, \object{PG 2112$+$059}, displayed long-term
variability decreasing by 0.18 magnitudes in the V-band over a period
of almost four years.

The long term variability of large samples of optically-selected
quasars have been studied over decades (Hook et al. 1994; Cristiani et
al. 1996). In these samples, a strong negative correlation between
variability and quasar luminosity was found with the more luminous
quasars displaying less variability. This result is interesting
considering that RQQSOs in this sample all have high luminosities
($-27<M_\mathrm{V}<-30$), and no rapid optical or night to night
variability was detected in any of the sources.

\subsection{Radio results on RQQSOs}

The radio spectra of RQQSOs probably have contributions from three
components: (i) optically thin synchro\-tron from star forming regions
in the disk of the host galaxy and in a circumnuclear starburst, (ii)
optically thin synchrotron from an extended or possibly jet-like
component powered by an active nucleus and (iii) partially opaque
synchrotron from a compact VLBI-scale core. In an extensive survey of
RQQSOs, Kellermann et al. (1994) obtained VLA maps which show extended
and double lobe radio structures in some sources that are similar to
those observed in radio-loud quasars. The RQQSOs mostly have a radio
luminosity well in excess of the $10^{22}~\mbox{W Hz}^{-1}$ found for
most normal spiral and elliptical galaxies and hence the radio
emission is not simply that from the underlying galaxy. It is quite
possible that for sensitivity reasons quasars with additional
low-surface brightness features may have been missed in the VLA
mapping of the radio-quiet sources. Nevertheless, a number of the
radio-quiet quasars in the Kellermann et al. (1994) survey (e.g.
\object{0953$+$41}; \object{1116$+$21}; \object{1634$+$70}) are well
resolved and show extended structure ranging between 49 and
$\sim300~\mbox{kpc}$. Recent VLA results (Barvainis et al. 1996)
revealed heterogenous spectral shapes in the radio spectra of a sample
of RQQSOs that could be classified into general categories similar to
radio loud quasars. Furthermore variability was discovered for seven
sources most of which had flat or inverted radio spectra. In one
source VLBI revealed that essentially all the flux emanated from one
compact sub-parsec core. The radio results on these types of sources
appear to be inconsistent with starburst models and imply that the
cores of many RQQSOs may be scaled down versions of those found in
radio loud quasars (Stein 1996).

\subsection{The remarkable RQQSOs, \object{IRAS 13349$+$2438} and 
  \object{PG 1416$-$129}}

Recent new results have highlighted the unusual properties of two
RQQSOs, both of which lie near the top of the upper band of radio
emission for RQQSOs ($R\sim1$). The first source \object{IRAS
  13349$+$2438} was initially detected through its strong infrared
emission (Beichman et al. 1986) and has been classified as a
radio-quiet, infrared bright quasar with a value of $R=1.9$. It has a
redshift of $z=0.107$ and a high polarisation which rises from
1.4~\mbox{\%} at 2.2~\mbox{$\mu$m} (K-band) to 8~\mbox{\%} at
0.36~\mbox{$\mu$m} (U-band). Wills et al.  (1992) found no variability
of the polarisation or flux density on timescales from days to months
and discussed a bipolar geometry to account for its polarisation and
other properties. VLA observations revealed an unresolved source with
an unusual radio spectrum with a maximum flux density of 7~\mbox{mJy}
at a frequency of 5~\mbox{GHz}.  The origin of the peaked radio
emission is not understood but absorption of the radio emission may
occur in the dusty dense interstellar medium and also the
contributions of different spectra from several source components may
be involved.

\object{IRAS 13349$+$2438} has high polarisation, strong \ion{Fe}{ii}
emission and is radioquiet but no broad absorption lines (BAL) have
been observed.  \object{IRAS 13349$+$2438} has been observed on several
occasions by ROSAT where the source was found to vary by a factor of
4.1 in about one year and about 25~\mbox{\%} in one week (Brandt et
al.  1996) but showed no evidence for the large intrinsic absorption
of soft x-rays by cold neutral matter. The soft x-ray variability
excluded electron scattering for most of the soft x-rays and suggest
absorption by a warm ionized gas with internal dust.

Recent ASCA observations of \object{IRAS 13349$+$2438} discovered for
the first time rapid x-ray variability or blazar like behaviour in a
RQQSO. The source displayed intensity variations on two separate
occasions by factors of two on timescales of only a few hours without
any significant spectral changes (Brinkmann et al. 1996). There is
also some evidence in the x-ray data for even more rapid variability.
The 0.6 to 8~\mbox{keV} spectrum was fitted with a power law with
$\Gamma=2.40$ which is steeper than the average value of $\Gamma= 1.9$
found for RQQSOs in the ASCA band. Brinkmann et al (1996) point out
that the line of sight to the quasar may graze the edge of the torus
and suggest that small changes in viewing conditions could produce
marked changes in intensity and spectral shape.  However it is
difficult to understand how changes in viewing conditions could
produce such rapid variability with no significant spectral changes.
It is plausible that \object{IRAS 13349$+$2438} is an example of a radio
loud quasar that is viewed through a dusty ionizing outflow, possibly
associated with a merger, that severely attenuates the radio emission
so that the source is classified as a RQQSO. In this context it should
be noted that the quasar \object{III Zw 2} could be classified as a
RQQSO at 5~\mbox{GHz} but the radio spectrum rises steeply toward
higher frequencies and this source is radio-loud when the 90~\mbox{GHz}
flux density is adopted (Schnopper et al. 1978). Further observations
of \object{IRAS 13349$+$2438} across the full spectrum including VLBI
searches for a compact self-absorbed component or indeed multiple
source components may help elucidate the nature of this unusual hybrid
type source.

The second unusual RQQSO is \object{PG 1416$-$129} at $z=0.129$. It
has been classified (Turnshek \& Grillmair 1986; Ulrich 1988) as a
broad absorption line quasar \linebreak (BALQSO). The value of $R$ is
1.1 and similar to \object{IRAS 13349$+$2438} it lies near the top of
the upper band for RQQSOs which is also heavily populated with BALQSOs
(Francis et al. 1995). This source has a soft x-ray excess (de Kool \&
Meurs 1994) and the hardest spectral index of any source in the energy
range 2--20~\mbox{keV} when observed with GINGA (Williams et al.
1992). Staubert \& Maisack (1996) detected this bright RQQSO at
energies above 50~\mbox{keV} with the OSSE telescope on CGRO and the
flux was found to be variable on a timescale of days during the 14 day
observation.

The BAL classification of this source has been questioned by Green \&
Mathur (1996) who proposed that large values of the optical to X-ray
slope ($\alpha_\mathrm{ox}>1.8$) be the defining characteristic of
BALQSOs. They report a low value of ($\alpha_\mathrm{ox}\sim1.4$) for
\object{PG 1416$-$129} and suggest further observations with HST to
check the BAL classification. VLA observations of \object{PG 1416$-$129}
reveal unusual results for a RQQSO: (i) the radio source consists of
an unresolved core coincident with an extended component that is
assumed not to be an unrelated background source (Kellermann et al.
(1994), (ii) the source varied by a factor of 4.5 over a period of ten
years and could have been more variable given the very limited
monitoring at radio frequencies and (iii) the radio spectrum is
consistent with a flat or inverted spectrum (Barvainis et al. 1996).
No VLBI observations have been reported for \object{PG 1416$-$129} but
one of the other variable RQQSOs \object{PG 1216$+$069} was detected
with essentially all of its flux in the VLBI core and the high
brightness temperature limit confirmed the self-absorbed synchrotron
hypothesis for the flat spectrum component (Barvainis et al. 1996).
This component may well be a scaled-down version of the radio sources
observed in radio loud quasars and future VLBI observations of
\object{PG 1416$-$129} may also reveal a similar component.

\section{Conclusions}

A long-term survey of a sample of high luminosity
($-27<M_\mathrm{V}<-30$) and medium to high redshift ($0.466<z<4.7$)
radio-quiet quasars was undertaken, in order to search for short and
long term optical variability on timescales of hours to years. A large
sample of 23 RQQSOs was observed, with a total observation time of
about 60 hours spread over a period of several years. Long-term
variability was detected in the RQQSO \object{PG 2112$+$059} when it
varied by 0.18 magnitudes in the V-band between 1992 and 1996. No
rapid variability was observed in any of the sources in this sample of
RQQSOs. The finding charts are included because they identify the
RQQSO and the reference stars used in the photometry and hence are
available to other observers.

There have been a few reports of rapid optical variability in a number
of RQQSOs but these reports are not in conflict with the results
presented here because such small variability would not have been
detected in many of the sources monitored in this survey. The unusual
properties of two sources are highlighted. These sources were not
monitored in this survey but have recently been added to the list of
sources for study. The remarkable source \object{IRAS 13349$+$2438}
combines some of the properties of blazars and radio quiet quasars and
hence further observations may elucidate the nature of this hybrid
source. The two unusual sources have $R$ values near the top of the
range for RQQSOs and also have unusual radio spectra that may signify
the presence of several source components. Further observations with
VLA and VLBI should reveal new and enlightening views on the radio
properties of these sources.

\begin{acknowledgements}

  The Jacobus Kapteyn Telescope on the island of La Palma is operated
  by the Royal Greenwich Observatory at the Spanish Observatorio del
  Roque de los Muchachos of the Institato de Astrofisica de Canarias.

  We are grateful to Catherine Handley and Matt Delaney for
  their help in the preparation of this paper.
 
\end{acknowledgements}



\begin{thebibliography}{}
 
 
\bibitem[1991]{abramowicz} Abramowicz M.A., Bao G., Lanza A., Zhang X.H., 
  1991, 
  A\&A 245, 454

\bibitem[1996]{barvainis} Barvainis R., Lonsdale C., Antonucci, 1996, 
  AJ 111, 1431

\bibitem[1986]{beichman} Beichman C.A., Soifer B.T., Helou G., et al., 1986, 
  ApJ 308, L1

\bibitem[1996]{brandt} Brandt W.N., Fabian A.C., Pounds K.A., 1996, 
  MNRAS 278, 326

\bibitem[1996]{Brinkmann} Brinkmann W., Kawai N., Ogasaka Y., Siebert J., 
  1996, 
  A\&A 316, L9

\bibitem[1996]{cristiani} Cristiani S., Trentini S., La Franca F., et al., 
  1996, 
  A\&A 306, 395

\bibitem[1988]{crusius} Crusius A., Schlickeiser R.A., 1988, 
  A\&A 196, 327

\bibitem[1994]{kool} de Kool M., Meurs E.J.A., 1994, 
  A\&A 281, L65

\bibitem[1995]{dermer} Dermer C.D., Gehrels N., 1995, 
  ApJ 447, 103

\bibitem[1984]{drew} Drew J. E., Boksenberg A., 1984, 
  MNRAS 211, 813

\bibitem[1993]{francis} Francis P.J., Hooper E.J.,  Impey C.D., 1993, 
  AJ 106, 417

\bibitem[1991]{gopal1} Gopal-Krishna, Subramanian K., 1991, 
  Nat 349, 766

\bibitem[1995]{gopal2} Gopal-Krishna, Sagar R., Wiita P.J., 1995, 
  MNRAS 274, 701

\bibitem[1993]{gopal3} Gopal-Krishna, Wiita P.J., Altieri B., 1993, 
  A\&A 271, 89

\bibitem[1980]{green1} Green R.F., Pier J.R., Schmidt M., et al., 1980, 
  ApJ 239, 483

\bibitem[1996]{green2} Green P.J., Mathur S., 1996, 
  ApJ 462, 637

\bibitem[1986]{hartig} Hartig G.F., Baldwin, J.A., 1986, 
  ApJ 302, 64

\bibitem[1979]{hazard1} Hazard C., Arp H.G., Morton D.C., 1979, 
  Nat 282, 271 

\bibitem[1984]{hazard2} Hazard C., Terlevich R., McMahon R., et al., 1984, 
  MNRAS 211, 45p

\bibitem[1987]{hewitt} Hewitt A., Burbidge G., 1987,  
  ApJS 63, 1

\bibitem[1994]{hook} Hook I.M., McMahon R.G., Boyle B.J., Irwin M.J., 1994,
  MNRAS 268, 305

\bibitem[1986]{hughes} Hughes P.A., Aller H.D., Aller M.F., 1986, 
  ApJ 341, 68

\bibitem[1991]{irwin} Irwin M.J., McMahon R.G., Hazard C., 1991, 
  in: The Space Distribution of Quasars, Crampton D. (ed.),
  APS Conference Series 21, p.\ 117

\bibitem[1995]{jang} Jang M., Miller H.R., 1995, 
  ApJ 452, 582

\bibitem[1989]{kellermann1} Kellermann K.I., Sramek R.A., Schmidt M., 
  Shaffer D.B., Green R., 1989, 
  AJ 98, 1195

\bibitem[1994]{kellermann2} Kellermann K.I., Sramek R.A., Schmidt M., 
  Green R., Shaffer D.B., 1994, 
  AJ 108, 1163

\bibitem[1981]{macalpine} MacAlpine G.M., Williams G.A., 1981, 
  ApJS 45, 113

\bibitem[1993]{mangalam} Mangalam A.V., Witta P.J., 1993, 
  ApJ 406, 420

\bibitem[1980]{marscher} Marscher A.P., 1980, 
  ApJ 239, 296 

\bibitem[1995]{mathur} Mathur S., Elvis M., Singh K.P., 1995, 
  ApJ 455, L9

\bibitem[1987]{mcbreen} McBreen B., Metcalfe L., 1987, 
  Nat 330, 348

\bibitem[1990]{miller} Miller L., Peacock J.A., Mead A.R.G., 1990, 
  MNRAS 244, 207

\bibitem[1984]{mitchell} Mitchell K.J., Warnock III A., Usher P.D., 1984, 
  ApJ 287, L3

\bibitem[1989]{neugebauer} Neugebauer G., Soifer B.T., Matthews K., 
  Elias J.H., 1989,
  AJ 97, 957

\bibitem[1988]{pica} Pica A.J., Smith A.G., Webb J.R., et al., 1988, 
  AJ 96, 1215

\bibitem[1991]{qian} Qian S.J., Quirrenbach A., Witzel A., et al. 1991, 
  A\&A 241, 15

\bibitem[1996]{rabbette} Rabbette M., McBreen B., Steel S., Smith N., 1996, 
  A\&A 310, 1

\bibitem[1995]{romero} Romero G.E., Surpi G., Vucetich H., 1995, 
  A\&A 301, 641 

\bibitem[1996]{sagar} Sagar R., Gopal-Krishna, Wiita P.J., 1996, 
  MNRAS 281, 1267

\bibitem[1978]{schnopper} Schnopper H.W., Delvaille J.P., Epstein J.P., 
  et al., 1978,
  ApJ 222, L91

\bibitem[1986]{smith} Smith E.P., Heckman T.M., Bothun G.D., Romanishin W., 
  Balick B., 1986, 
  ApJ 306, 64

\bibitem[1991]{sopp} Sopp H.M., Alexander, P., 1991, 
  MNRAS 251, 14p

\bibitem[1996]{staubert} Staubert R., Maisack M., 1996, 
  A\&A 305, L41

\bibitem[1996]{stein} Stein W.A., 1996, 
  AJ 112, 909

\bibitem[1996]{storrie} Storrie-Lombardi L.J., McMahon R.G., Irwin M.J., 1996, 
  ApJ 468, 121

\bibitem[1996]{taylor} Taylor G.L., Dunlop J.S., Hughes P.H., Robson E.I., 
  1996,
  MNRAS 283, 930

\bibitem[1992]{terlevich} Terlevich R., Tenorio-Tagle G., Franko J., 
  Melnick J., 1992, 
  MNRAS 255, 713

\bibitem[1986]{turnshek} Turnshek D.A., Grillmair C.J., 1986, 
  ApJ 310, L1

\bibitem[1988]{ulrich} Ulrich I.M.H., 1988, 
  MNRAS 230, 121

\bibitem[1981]{usher} Usher P.D., 1981, 
  ApJS 46, 117

\bibitem[1985]{veron1} V\'{e}ron-Cetty M.-P., V\'{e}ron P., 1985, 
  A Catalogue of Quasars and Active Galactic Nuclei, 2nd ed., Publ. ESO

\bibitem[1990]{veron2} V\'{e}ron-Cetty M.-P., Woltjer L., 1990, 
  A\&A 236, 69

\bibitem[1986]{wampler} Wampler E.J., 1986, 
  A\&A 161, 223

\bibitem[1988]{webb} Webb J.R., Smith A.G., Leacock R.J., et al., 1988, 
  AJ 95, 374

\bibitem[1992]{wiita} Wiita P.J., Miller H.R., Gupta N., 
  Chakrabarti S.K., 1992, 
  in: Variability of Blazars, Valtaoja E., Valtonen M. (eds.),
  Cambridge Univ. Press, Cambridge, p.\ 311

\bibitem[1992]{wills} Wills B.J., Wills D., Evans II N.J., et al., 1992, 
  ApJ 400, 96

\bibitem[1992]{williams} Williams O.R., Turner M.J.L., Stewart G.C., 
  Saxton R.D., 
  et al., 1992, 
  ApJ 389, 157

\bibitem[1995]{wilson} Wilson A.S., Colbert E.J.M., 1995, 
  ApJ 438, 62 

\end{thebibliography}
\end{document}